\pdfoutput=1

\documentclass[prl,aps,superscriptaddress,showpacs,reprint]{revtex4-1}
\usepackage{graphicx}
\usepackage{amsmath}
\usepackage{amssymb}
\begin{document}

\title{Charge tuning of non-resonant magneto-exciton phonon interactions in graphene} 

\author{Sebastian R\'{e}mi}
\affiliation{Boston University, Department of Physics, 590 Commonwealth Ave, Boston, MA 02215}

\author{Bennett B. Goldberg}

\affiliation{Boston University, Department of Physics, 590 Commonwealth Ave, Boston, MA 02215}
\affiliation{Boston University, Department of Electrical and Computer Engineering}
\affiliation{Boston University, Photonics Center, 8 St. Mary's St, Boston, MA 02215}

\author{Anna K. Swan}
 \email[]{swan@bu.edu}
\affiliation{Boston University, Department of Electrical and Computer Engineering}
\affiliation{Boston University, Department of Physics, 590 Commonwealth Ave, Boston, MA 02215}
\affiliation{Boston University, Photonics Center, 8 St. Mary's St, Boston, MA 02215}

\date{\today}

\begin{abstract}
Far from resonance, the coupling of the G-band phonon to magneto-excitons in single layer graphene displays kinks and splittings versus filling factor that are well described by Pauli blocking and unblocking of inter- and intra- Landau level transitions. We explore the non-resonant electron-phonon coupling by high-magnetic field Raman scattering while electrostatic tuning of the carrier density controls the filling factor. We show qualitative and quantitative agreement between spectra and a linearized model of electron-phonon interactions in magnetic fields. The splitting is caused by dichroism of left and right handed circular polarized light due to lifting of the G-band phonon degeneracy, and the piecewise linear slopes are caused by the linear occupancy of sequential Landau levels  versus $\nu$.
\end{abstract}


\maketitle 

When Dirac fermions in graphene are subjected to a perpendicular magnetic field $B$, the electronic states form discrete, degenerate Landau levels (LL) with energy of $E_{\pm, n}=\pm v_F\sqrt{2e\hbar Bn}$ ($n$ is the Landau level index) \cite{neto2009electronic,goerbig2011electronic}. Each level has spin(2) and valley(2) degeneracy with a system occupancy described by the filling factor $\nu = h\widetilde{n}/eB$, where $\widetilde{n}$ is the surface density of charges \cite{goerbig2007filling}.
At charge neutrality $\nu = 0$ and the $n=0$ level is half filled. A coherent superposition of inter-LL magneto-excitons and phonons can occur when the magnetic field is tuned so that an allowed LL transition is in resonance with the phonon energy \cite{ando2007magnetic,goerbig2007filling,goerbig2011electronic}. Systems where the electronic and phonon dephasing is smaller than the electron-phonon interaction strength show pronounced anticrossings of the energy of the phonon and magneto-exciton states \cite{ando2007magnetic,goerbig2007filling,goerbig2011electronic,PhysRevB.84.235138,kim2012magnetophonon,yan2010observation,kuhne2012polarization,faugeras2011magneto,yan2010observation}.
The first observation of strong electron-phonon coupling at resonance conditions was made by Raman spectroscopy of magneto-phonons in multi-layer graphene on SiC. However the sample was not pristine enough to exhibit coherent phonon magneto-exciton states\cite{faugeras2009tuning}. Coherent magneto-phonon Raman response has since been observed in graphite \cite{PhysRevB.84.235138,kim2012magnetophonon,yan2010observation} and decoupled surface layers of graphene on graphite crystals \cite{kuhne2012polarization,faugeras2011magneto,yan2010observation}. Recently, magneto-phonon resonances have also been observed on single layer graphene exfoliated on SiO$_2$ \cite{PhysRevLett.110.227402,kossacki2012circular}.

Magneto-exciton transitions couple to orthogonal states of the degenerate modes of the G-band phonon. Symmetry allowed transitions obey the selection rule $\Delta\left|n\right|=\pm 1$ \cite{PhysRevB.84.235138}. Experimentally selective excitation of the orthogonal states is achieved using cross circular polarized optical excitation and detection channels. In a $\sigma^+/\sigma^-$ configuration, only states coupling to $\Delta n=+1$ transitions can be observed, while only states coupling to $\Delta n=-1$ transitions are observed in $\sigma^-/\sigma^+$ polarization configuration \cite{kossacki2012circular,PhysRevLett.110.227402,PhysRevB.84.235138,kuhne2012polarization}.
These selection rules create optical dichroism for doped graphene \cite{goerbig2007filling,ando2007magnetic} and have been observed in graphene on SiO$_2$ due to partial Pauli blocking of the initial or final Landau level states \cite{PhysRevLett.110.227402,kossacki2012circular}. 
High quality graphene on SiC or on graphite can be considered charge neutral, while exfoliated graphene on SiO$_2$ typically shows accidental doping from sample preparation and impurities in the substrate. So far, only limited control of the doping level $\widetilde{n}$ has been achieved by annealing and gas exposure in between scanned B-field measurements \cite{PhysRevLett.110.227402}. However, for constant $\widetilde{n}$ the filling factor $\nu$ varies with the B-field strength. We use graphene field effect devices, widely used in transport \cite{novoselov2004electric,novoselov2005two,zhang2005experimental} as well as Raman measurements \cite{pisana2007breakdown,yan2007electric,stampfer2007raman} to provide full control of the charge density and decouple the effects of magnetic field and filling factor dependence.

Here we show for the first time charge carrier density dependent magneto Raman measurements on single layer graphene field effect devices at constant magnetic fields. Contrary to previous magneto-phonon studies, we explore the Raman response in a magnetic field far from magneto-phonon resonance conditions. We predict a linear dependence with $\nu$ for the non-resonant regime, rather than the $\sqrt{\nu}$ behavior predicted for the on-resonance response. Notably, at constant magnetic field we observe pronounced splittings and slope changes of the G-band phonon energy as a function of $\nu$, the LL filling. We show that the structure in the electron-phonon coupling is due to the occupancy of Landau level magneto-exciton transitions coupled to the phonon. In contrast to on-resonance measurements, no single transition dominates the coupling, and several inter and intra-band transitions have to be considered to account for the experimental observations. Splitting is due to the different filling factor dependent response to left and right hand polarized light.   

For control of the charge carrier density we fabricated field effect devices based on single layer graphene where the charge density is controlled by the gate voltage. Graphene is exfoliated onto a 300 nm thick SiO$_2$/Si substrate and the devices are fabricated using standard microfabrication processes. We confirm the single layer character by Raman spectroscopy and measurement of the optical contrast \cite{ferrari2006raman,casiraghi2007rayleigh,ni2007graphene,blake2007making}.

\begin{figure}
\includegraphics{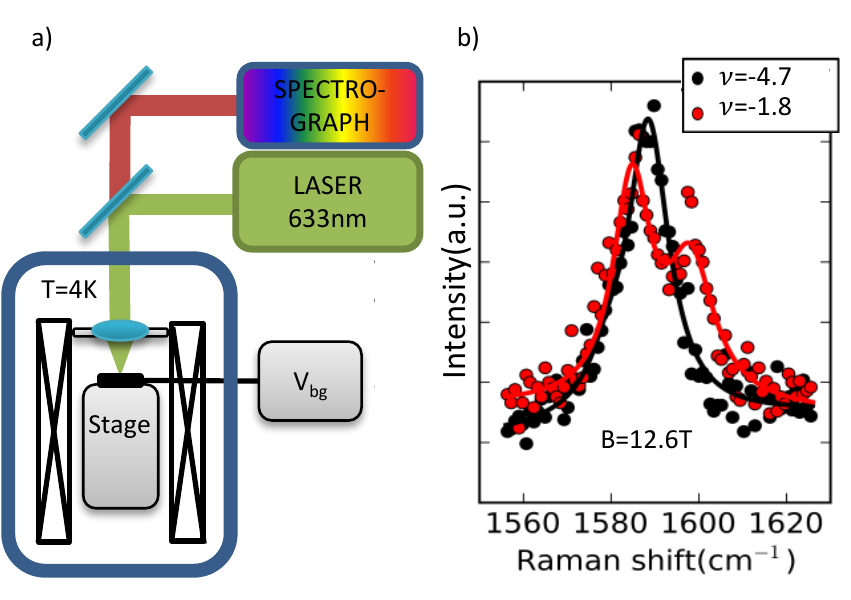}
\caption{\label{setup}a) Schematic view of the cryogenic Raman setup using linearly polarized light. b) G-band Raman spectra at $B=12.6T$. Black spectrum measured at $\nu=-4.7$ ($V_{bg}=-20\: V$). The red spectrum shows visible splitting of the G-band at $\nu=-1.8$ ($V_{bg}=-8\: V$). Solid lines are double-peaked Lorentzian fits.}
\end{figure}

Raman measurements are performed at 4K in a He bath cryostat, schematically drawn in Fig.\,\ref{setup}(a). The sample is mounted on a piezoelectric x-y-z stage at the focus of a confocal, free-space microscope cooled with He exchange gas and placed in the center of a superconducting magnet with an accessible range of $B=\left\lbrace 0\: T,12.6\: T \right\rbrace$.
The Raman response is excited using a HeNe Laser at $\lambda=632.8\: nm$ with a diffraction limited spot size $\sim 1\: \mu m$. The excitation laser is linearly polarized, although we do not monitor or optimize the polarization. Since our experiment uses linear polarized light without an analyzer, we measure both  $\sigma^+/\sigma^-$  and $\sigma^-/\sigma^+$ transitions, i.e. both $\Delta |n|=\pm1$ transitions are detected simultaneously. Scattered light is filtered by a long pass filter to remove the laser light, collected by a single mode fiber and analyzed using a conventional grating spectrometer.

Fig.\,\ref{setup}(b) illustrates the effect of applying a backgate voltage $V_{bg}$ at finite magnetic field of $B=12.6\: T$. For $\nu =-4.7$ $(V_{bg}=-20\: V)$ the G-band is symmetric with a single peak (black line), but splits into 2 peaks for $\nu=-1.8$ $(V_{bg}=-8\: V)$ due to the interaction of the optical phonons with the discrete Landau levels \cite{kias2009} .

Characterization of the phonon response versus charge density was performed by Raman spectroscopy while sweeping the backgate in the range $V_{bg}=\left\lbrace -40\: V,40\: V \right\rbrace$ both at $B=0\: T$ as well as $B=12.6\: T$.
\begin{figure}[!hb]
\includegraphics[width=\columnwidth]{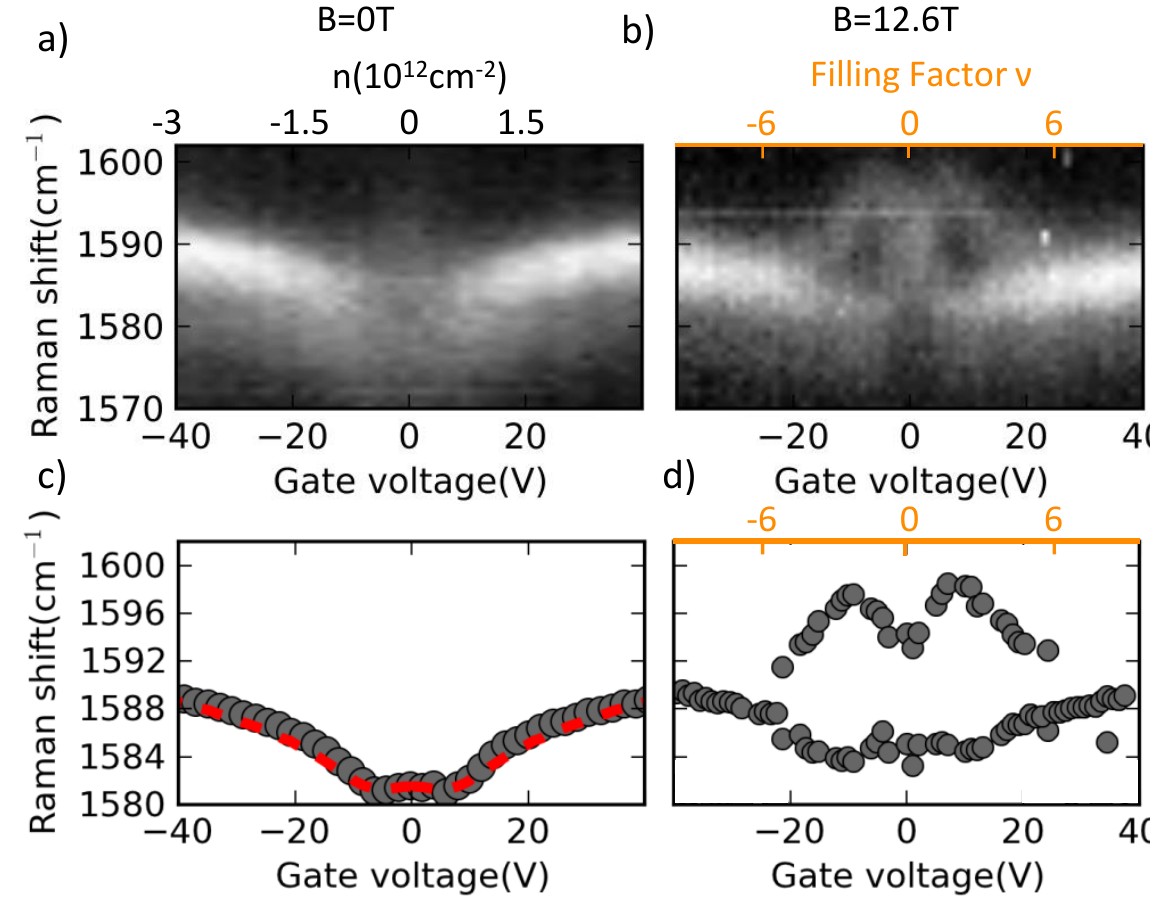}
\caption{\label{bcompare}
Raman intensity map of the G-band Raman spectra as a function of applied backgate voltage. The charge density and filling factor are indicated on top of the figures. a) Measurements for $B=0\: T$. b) $B=12.6\: T$. A clear splitting is visible for voltages $\left|V\right|\leq 20\: V$. c) and d)  show phonon energies from Lorentzian fits in a) and b). Dashed red line is a fit using a model of the phonon anomaly in graphene \cite{tsuneya2006anomaly}.}
\end{figure}
Fig.\,\ref{bcompare} shows a Raman intensity map of the observed spectra as a function of $V_{bg}$. We extract the position of the G-band by fitting the spectra with single ($B=0\: T$) and double lorentzian ($B=12.6\: T$) functions shown in Fig.\,\ref{bcompare}(c) and (d).

Measurements results at $B=0\: T$ are shown in Fig.\,\ref{bcompare}(a) and (c).
The behavior of the G-band  as a function of charge density has previously been studied theoretically and experimentally \cite{pisana2007breakdown,yan2007electric,stampfer2007raman,tsuneya2006anomaly}.  We use the model of \cite{tsuneya2006anomaly} to fit the data, and also include the effects of inhomogeneous broadening due to charge carrier density fluctuations \cite{yan2007electric}. In Fig.\,\ref{bcompare}(c) we plot the fit results (red dashed line) as a function of $V_{bg}$, and extract the following system parameters: The electron-phonon coupling strength $\lambda = 4.8\times 10^{-3}$, the phenomenological broadening parameter \cite{tsuneya2006anomaly} $\delta=10\:  meV$, the unperturbed phonon energy at $B=0\: T$, $\varepsilon=1582.0\: cm^{-1}$, the inhomogeneous broadening $\delta \widetilde{n}=0.3\times 10^{12}\: cm^{-2}$ (standard deviation of a Gaussian distribution) and finally the Fermi velocity $v_F=1.10\times 10^6\: ms^{-1}$.  For more details on the qualitative and quantitative description at $B=0\: T$ we refer to the supplementary material.

Fig.\,\ref{bcompare}(b) and (d) shows a very different behavior for $B=12.6\: T$.
G-band splitting starts around $\nu$=$-6$ $(V_{bg}$\,$\approx$\,$-25\: V)$ then reaches a maximum at $\nu$=$-2$ $(V_{bg}\approx-9\: V)$, disappears at $\nu$=$0$ $(V_{bg} = 0\: V)$ and repeats symmetrically for $\nu>0$. The largest magnitude of the splitting is $\sim 12\: cm^{-1}$. At $B=12.6\: T$, the nearest magneto-phonon resonances are the transitions between $\left|n\right|=0$ and $\left|n\right|=1$ at $B=25\: T$ and $\left|n\right|=1$ and $\left|n\right|=2$ at $B=4.1\: T$.

\begin{figure*}
\includegraphics{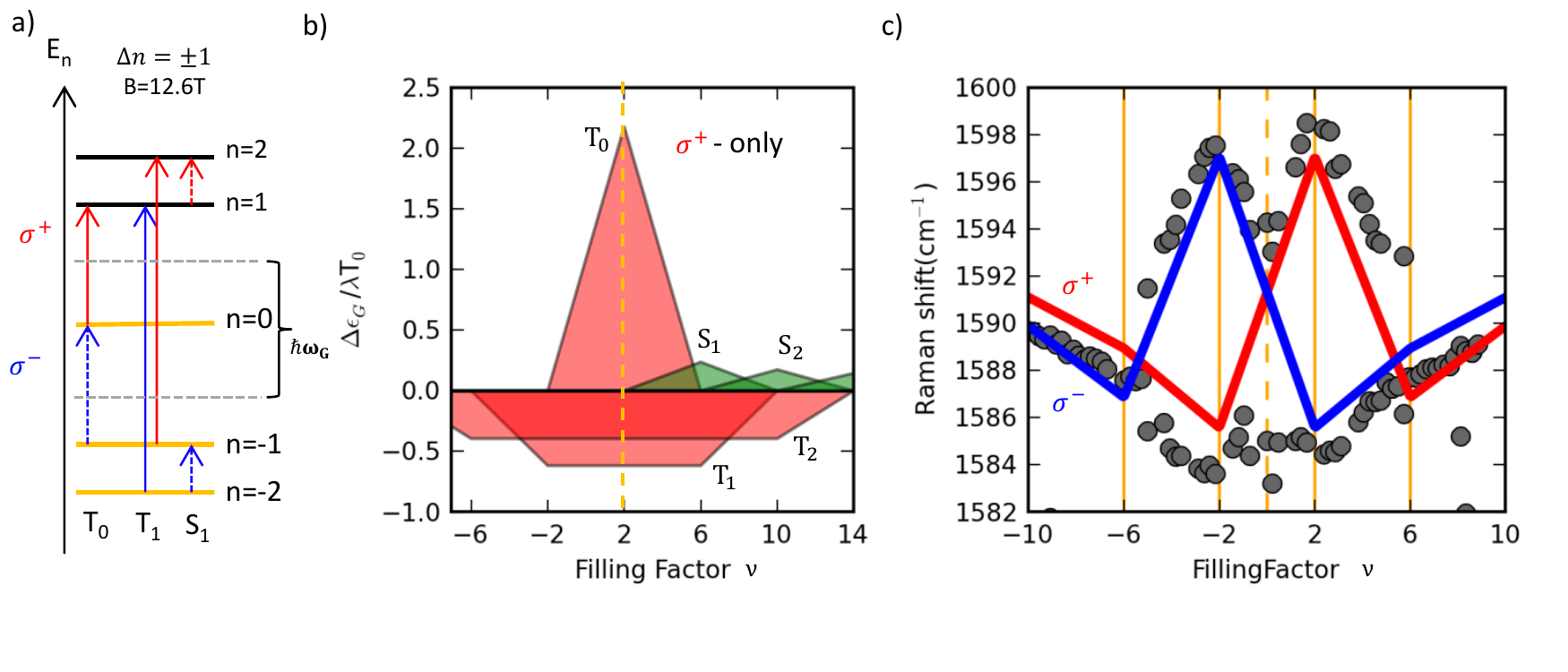}
\caption{\label{bfield} a) Schematic view of the Landau level spectrum at $B=12.6\: T$, filling factor $\nu=2$ and the lowest Landau level transitions participating in magneto-phonon coupling. Filled electronic states are highlighted using orange color. Red and blue arrows show transitions allowed by the selection rule $\Delta\left|n\right|= \pm 1$. Dashed arrows mark Pauli blocked transitions. Circular arrows represent the angular momentum involved in the transitions b) Relative strength and filling factor dependence of individual terms of the phonon self energy for $\sigma^+$-transitions. Terms describing interband transitions are shaded red, intraband transitions are shaded green.  c) Phonon energy as a function of the filling factor at $B=12.6\: T$. Vertical orange lines mark specific filling factors at $\nu=-6,-2,0,2,6$ where the n=-1,0,1 levels are completely filled/depleted with charge carriers ($\nu=0$ corresponds to half filling of $n$=$0$ level). The calculated magneto-phonon energies according to Eqn.\,(\ref{greens}) are plotted as solid red($\Delta n = +1$) and solid blue($\Delta n = -1$) lines.}
\end{figure*}

Following \cite{goerbig2007filling,ando2007magnetic,kossacki2012circular} we consider the phonon energy $\varepsilon_{\mathcal{A}}$ in Eqn.\,(\ref{greens-full}). The index $\mathcal{A}$ denotes the two orthogonal circularly polarized phonon states accessed by $\sigma^+$ or $\sigma^-$ circularly polarized light.

\begin{eqnarray}
\label{greens-full}
\varepsilon_{\mathcal{A}}^2-\varepsilon_0^2 &=& 2\varepsilon_0 \lambda T_0^2 \left[ \sum_{n=0}^N\left(\frac{f_{\mathcal{A},n}\left(\nu\right) T_n}{\left(\varepsilon_{\mathcal{A}}+i\delta\right)^2-T_n^2}-\frac{1}{T_n}\right)\right.\nonumber \\
& &\left. +\sum_{n=1}^{N}\frac{f_{\mathcal{A},n}\left(\nu\right) S_n}{\left(\varepsilon_{\mathcal{A}}+i\delta\right)^2-S_n^2}\right]
\end{eqnarray}
Here $\varepsilon_0$ is the unperturbed phonon energy, $\lambda$ is the dimensionless electron-phonon coupling parameter and $\delta$ is the phenomenological broadening introduced by Ando \cite{ando2007magnetic}. The two sums are the contribution from all the interband and intraband asymmetric transitions (see Fig.\,\ref{bfield}(a)). The interband transitions have energy $T_n=T_0\left(\sqrt{n+1}+\sqrt{n}\right)$ and intraband transitions $S_n=T_0\left(\sqrt{n+1}-\sqrt{n}\right)$ where $T_0=v_F\sqrt{2e\hbar B}$. For $B=12.6\: T$, $T_0 \approx 140\: meV$ compared with $\sim 196\: meV$ for the G band phonon.
The filling factor dependence of the coupling to the highly degenerate Landau level states is described by the factor $f_{\mathcal{A},n}$. For interband transitions $f_{\mathcal{A},n}$ is defined by
\begin{eqnarray}
\label{fterm}
f_{\sigma^+,n}&=&(1+\delta_{n,0})(\bar{\nu}_{-n}-\bar{\nu}_{+(n+1)})\nonumber\\
f_{\sigma^-,n}&=&(1+\delta_{n,0})(\bar{\nu}_{-(n+1)}-\bar{\nu}_{+n})
\end{eqnarray}
Here  $\bar{\nu}_{n} $ is the normalized  filling factor describing the fraction of filling of the $n$’th Landau level. It is related to the filling factor $\nu$ by $\bar{\nu}_{\pm n} = \left[\nu -(4(\pm n)-2)\right]/4$ since each Landau level state is fourfold degenerate.
Hence $0<\bar{\nu}_{n}<1$. The definition of $f_{\mathcal{A},n}$ for intraband transition is easily obtained by replacing the index $\mp n$ by $\pm n$. In  Fig.\,\ref{bfield}(a) we illustrate the level occupation for a filling factor $\nu=2$, i.e. the $n=0$ level is completely filled (partial filling factor $\bar{\nu}_0=1$). Hence,  the transition between $n=-1$ and $ n=0$ is Pauli blocked (dashed lines in figure) and  $f_{\sigma^-,0}=0$, while all transitions originating from the $n$=$0$ level have maximum strength due to the high density of occupied states that can be excited, with $f_{\sigma^+,0}=1$.

Eqn.\,(\ref{greens-full}) is valid for all $B$ fields and charge states, although it is cumbersome to use. Near resonance, a single resonant term dominates, so other terms can be neglected in solving Eqn.\,(\ref{greens-full}). The solution is described by a two-level coupled mode model \cite{yan2010observation,PhysRevLett.110.227402}
\begin{equation}
\label{resonant-shift}
\varepsilon_{\mathcal{A}}^{\pm}=\frac{T_n+\varepsilon_0}{2}\pm\sqrt{\left(\frac{T_n-\varepsilon_0}{2}\right)^2+\frac{\lambda T_0^2}{2}f_{\mathcal{A},n}}
\end{equation}
Eqn.\,(\ref{resonant-shift}) describes the anticrossing between the coherent coupled states $\varepsilon_{\mathcal{A}}^{\pm}$. The index $\pm$ refers to the upper and lower branches of the anticrossing.

In the non-resonant regime, where our experiment is performed, the approximation leading to Eqn.\,(\ref{resonant-shift}) is not valid. Since  $\Delta\varepsilon_{\mathcal{A}} = \varepsilon_{\mathcal{A}} - \varepsilon_0$ is small, Eqn.\,(\ref{greens-full}) can be linearized by noting that $\varepsilon_{\mathcal{A}}^2-\varepsilon_0^2 \approx (\varepsilon_{\mathcal{A}}-\varepsilon_0)2\varepsilon_0$, and  replacing the $\varepsilon_{\mathcal{A}}$ in the denominator by the unperturbed phonon frequency $\varepsilon_0$ \cite{ando2007magnetic}. The shift of the G-band in the non-resonant regime is then given by

\begin{eqnarray}
\label{greens}
\Delta\varepsilon_{\mathcal{A}} &=& \mathrm{Re}\left\lbrace \lambda T_0^2 \left[ \sum_{n=0}^N\left(\frac{f_{\mathcal{A},n}\left(\nu\right) T_n}{\left(\varepsilon_{0}+i\delta\right)^2-T_n^2}-\frac{1}{T_n}\right)\right.\right.\nonumber \\
& &\left.\left. +\sum_{n=1}^{N}\frac{f_{\mathcal{A},n}\left(\nu\right) S_n}{\left(\varepsilon_{0}+i\delta\right)^2-S_n^2}\right]\right\rbrace
\end{eqnarray}

The expressions Eqn.\,(\ref{resonant-shift}) and Eqn.\,(\ref{greens}) are distinguished by  the numbers of terms needed, and by their different  filling factor dependence. In the resonance approximation Eqn.\,(\ref{resonant-shift}) the shift is proportional to $ \sqrt{\nu}$ while in the non-resonant case Eqn.\,(\ref{greens}) is linear in the filling factor $\nu$.

We now consider the effect of the charge tuning in Eqn.\,(\ref{greens}) with fixed $B$ field. 
In order to evaluate the contributions from participating inter- and intra-band transitions, we calculate the individual contribution to the phonon energy shift $\Delta\varepsilon_n$ from each term in Eqn.\,(\ref{greens}) evaluated at $B=12.6\: T$. Shown in Fig.\,\ref{bfield}(b) are the $\Delta\varepsilon_n$ for the lowest lying inter- and intra-band transitions for $\sigma^+$ polarization, i.e. transitions with $\Delta n =+1$. We use the values for $\varepsilon_0$, $\lambda$, $v_F$ and $\delta$ from the $B=0\: T$ fit. Curves are labeled as T$_n$ (red shaded) for inter-band and S$_n$ (green shaded) for intra-band transitions, normalized by $\lambda$T$_0$. The largest contribution is due to the T$_0$ term which shows a strong peak at $\nu=2$. The contributions from the remaining inter-band transitions are strong as well. The shift due to the T$_1$ term is 28\% and the T$_2$ term is still at 17.9\% relative to the shift caused by the T$_0$ term.
The action of intra-band terms are restricted to a smaller range of $\nu$ values, but their strength can be significant nevertheless (S$_1$/T$_0$ $\approx$ 10\%). The case of $\sigma^-$ polarization is completely symmetric relative to the point $\nu = 0$.

Finally, we combine these contributions and compare to our experimental data (Fig.\,\ref{bfield}(c)). We only include the first 5 terms of Eqn.\,(\ref{greens}) in our calculation, since the neglected terms only cause a small overall downshift of $\sim -1cm^{-1}$ in the range of our measurement (See supplementary material). The phonon energy is plotted versus filling factor rather than charge density to highlight the correspondence between the filled Landau levels and the extrema of kinks in the slope (orange lines). The solid red and blue lines are the numerical results for the energies $\varepsilon_{\sigma^+}$ and $\varepsilon_{\sigma^-}$. We emphasize that no adjustable parameters have been used to compare data and theory at $B=12.6\: T$ -  all parameters are determined from the $B=0$ data.
The splitting between $\varepsilon_{\sigma^+}$ and $\varepsilon_{\sigma^-}$ is maximal at $\nu=-2$ and $\nu=2$ where the coupling strength to the T$_0$ transitions corresponding to $\Delta|n|=\pm1$ respectively are strongest. Fig.\,\ref{bfield}(b) also explains the kink in slope $\nu=\pm6$ due to Pauli blocking of T$_0$ and T$_1$ transitions. The upshift with increasing $|\nu|$ is caused by the linear decrease of the T$_1$ transition as the LL $n=\pm1 $ are filled or emptied, respectively. In principle scanning to larger absolute value $|\nu|$ will reveal higher and higher $n$ transitions. 
There is some evidence in the data of a small contribution  from the symmetric transitions $\Delta n = 0$  \cite{PhysRevB.84.235138} at the same reduced coupling strength seen by \cite{faugeras2011magneto,kuhne2012polarization}, however, it is not conclusive (see supplementary material). We could not resolve the small splitting of $\sim 1\: cm^{-1}$ between $\varepsilon_{\sigma^+}$ and $\varepsilon_{\sigma^-}$ for $\left|\nu\right|>6$ predicted by the model.

The extracted values of Fermi velocity  $v_F=1.1\times 10^6\: ms^{-1}$ and electron phonon coupling $\lambda=4.8\times 10^{-3}$ agree well with those determined in previous experiments on SiO$_2$ \cite{PhysRevLett.110.227402,kossacki2012circular}. However, for graphene on graphite, a lower $v_F$,  and a $23\%$ higher value of  $\lambda$ is reported \cite{yan2010observation}. (See supplementary material for further discussion).

By focusing on the non-resonant regime we have discovered fine structure in the G-band optical phonon in single layer graphene at high magnetic fields as a function of charge density. The observed behavior is caused by coupling between the phonon and magneto-exciton far from resonance that results in a linear dependence on filling factor, in contrast to on-resonant coupling that leads to a square root dependence on filling factor. High magnetic field Raman scattering with electrostatic tuning of the charge carrier density allows us to explore the filling factor dependent coupling strength of orthogonal, non-resonant magneto-phonon states as they are being turned on and off. By including coupling to many Landau level transitions we show qualitative and quantitative agreement with numerical calculations of a linearized model of electron-phonon interactions in magnetic fields. The measured coupling strength, broadening and Fermi velocity is in good agreement with independent observations at $B=0\: T$ and earlier experiments.
\\

\begin{acknowledgments}
We thank Mengkun Liu for help with sample preparation and Mark Goerbig and Alex Kitt for discussions.
\end{acknowledgments}

\bibliography{magnetophonon}

\end{document}